\documentstyle[a4,epsf]{article}
\title{\bf The Higgs mass bound \\ in the SUSY multi-Higgs-doublet model}
\author{{\bf Yutaka Sakamura}\footnote{Email: sakamura@th.phys.titech.ac.jp} \\
{\footnotesize\it Department of Physics, Tokyo Institute of Technology} \\
{\footnotesize\it Oh-okayama, Meguro, Tokyo 152-8551, Japan}}
\date{}
\newcommand{\bfQ}{\mbox{\boldmath $Q^{3}$}}
\newcommand{\bfU}{\mbox{\boldmath $U^{3c}$}}
\newcommand{\bfD}{\mbox{\boldmath $D^{3c}$}}
\newcommand{\bfHdi}{\mbox{\boldmath $H_{2i-1}$}}
\newcommand{\bfHui}{\mbox{\boldmath $H_{2i}$}}
\newcommand{\bfHuj}{\mbox{\boldmath $H_{2j}$}}
\newcommand{\ovmt}{\overline{m}_{t}}
\newcommand{\ovmb}{\overline{m}_{b}}
\newcommand{\ovmz}{\overline{m}_{Z}}
\newcommand{\mx}{\mbox{\scriptsize max}}
\begin{document}
\maketitle
\begin{abstract}
The upper bound on the mass of the lightest Higgs boson is provided 
in the supersymmetric standard model with multi-Higgs doublets, 
up to two-loop order. 
Relatively large corrections are expected from the experimentally
unconstrained extra Yukawa couplings.
We calculate it including the two-loop leading-log contributions 
as a function of 
the scale $\Lambda$, below which the theory remains perturbative.
Even for $\Lambda=10^{4}$~GeV, we obtain the upper bound of 140~GeV,
which is heavier than that of the MSSM by 10~GeV.
\end{abstract}

\section{Introduction}
There is a famous upper bound on the mass of the lightest Higgs boson 
in the minimal supersymmetric standard model (MSSM), {\it i.e.}, 
which is lighter than Z boson at tree level \cite{Ino;Kaku}.
This comes from the fact that the quartic couplings of the neutral Higgses are 
determined by only the weak gauge couplings $g$ and $g'$ 
due to the supersymmetry (SUSY).
However, since the top Yukawa coupling is large, this tree-level upper bound 
receives large corrections \cite{Okada;Yama,Ellis;R;Z,Haber;Hemp,Okada;Yama-2}.
Further, QCD correction, which begins from a two-loop order, is also large 
\cite{Kodaira,Hemp;Hoang,Casas-2,3H,Hein;Holl,Zhang}, because
the strong gauge coupling $\alpha_{s}$ first appears at two-loop order, 
in other words, two-loop is the lowest order in terms of $\alpha_{s}$.
Consequently, the upper bound can be set about 130~GeV in the MSSM 
when all the squark masses are at order of 1~TeV.

Here we will provide the upper bound on the mass of the lightest Higgs boson
in the supersymmetric standard model with arbitrary number of Higgs doublets,
which is the minimal extension of the MSSM in the meaning that 
its Higgs sector is composed of doublets only.
This model has the same upper bound as the MSSM up to one-loop level.
So, we must include at least two-loop corrections to the upper bound 
in order to see the difference from the MSSM.
Actually, the two-loop order corrections include the contributions of 
the extra Yukawa couplings, which are characteristic of 
the multi-Higgs-doublet model and can be relatively large.

We demand that the theory remains perturbative up to the scale 
$\Lambda$ in order to set the upper bound on the extra Yukawa
couplings, which are unconstrained experimentally.
Then the resulting upper bound $m_{h}$ depends on the scale $\Lambda$, and
becomes more stringent as $\Lambda$ is larger.
For example, we obtain $m_{h}=128$~GeV in the case of $\Lambda=10^{16}$~GeV, 
which is almost the same as the upper bound in the MSSM.
On the other hand, for $\Lambda=10^{4}$~GeV, which is one order above 
the SUSY-breaking scale $m_{S}$, the upper bound is lifted up by 10~GeV due to
the large extra Yukawa couplings.
Throughout this letter, the soft SUSY-breaking squark masses are assumed 
to be 1~TeV and degenerate for simplicity.

\section{Up to one-loop order}
We denote the number of the Higgs doublets as $2n$.\footnote{The number of the
Higgs doublets must be even because of the anomaly cancellation.}
Then the superpotential of this model is written as
\begin{eqnarray}
 W&=&\sum_{i=1}^{n}h_{2i}\bfQ\bfHui\bfU-\sum_{i=1}^{n}h_{2i-1}\bfQ\bfHdi\bfD
 \nonumber \\
 &&+\sum_{i,j=1}^{n}\mu_{2i-1,2j}\bfHdi\bfHuj+h.c.+\cdots,
\end{eqnarray}
where {\boldmath $Q^{3}$, $U^{3c}$} and {\boldmath $D^{3c}$} are 
the superfields of the left-handed quark doublet in the third generation, 
right-handed top quark and right-handed bottom quark respectively, and
$\bfHdi$ and $\bfHui$ are the superfields of the Higgs doublets.
The ellipsis denotes the terms involving quark superfields in the first and
second generations and lepton superfields, which are irrelevant to our result.

With general soft SUSY-breaking terms, the tree-level Higgs potential of 
this model is given by
\begin{eqnarray}
 V^{(0)}(\phi^{0}_{1},\cdots,\phi^{0}_{2n},\phi^{0*}_{1},\cdots,\phi^{0*}_{2n})
 & = & \frac{g^{2}+g'^{2}}{8}(\phi^{0\dagger}_{d}\phi^{0}_{d}-\phi^{0\dagger}
 _{u}\phi^{0}_{u})^{2} \nonumber \\
 &&+\phi^{0\dagger}_{d}M_{d}^{2}\phi^{0}_{d}+\phi^{0\dagger}_{u}M_{u}^{2}
 \phi^{0}_{u}-(\mbox{}^{t}\phi^{0}_{d}M_{du}^{2}\phi^{0}_{u}+h.c.),\nonumber\\
\end{eqnarray}
where
\begin{equation}
\phi^{0}_{d}\equiv\mbox{}^{t}(\phi^{0}_{1},\phi^{0}_{3},\cdots,\phi^{0}_{2n-1})
\hspace{5mm},\hspace{5mm}
\phi^{0}_{u}\equiv\mbox{}^{t}(\phi^{0}_{2},\phi^{0}_{4},\cdots,\phi^{0}_{2n}),
\end{equation}
are the neutral components of the Higgs doublets, which couple to the down-type
quarks and to the up-type quarks respectively.
The matrices $M_{u}^{2}$ and $M_{d}^{2}$ are $n\times n$ Hermitian matrices 
and $M_{du}^{2}$ is generally an $n\times n$ complex matrices.  
The couplings $g$ and $g'$ are the gauge couplings of 
$SU(2)_{L}$ and $U(1)_{Y}$ respectively.
We concentrate our attention only on the neutral Higgs fields 
since only neutral fields may take VEVs after the breakdown of the electroweak
symmetry.

Now we can rotate the Higgs fields so that 
\begin{equation}
 \langle\phi^{0\prime}_{1}\rangle=\frac{v_{d}}{\sqrt{2}},\;
 \langle\phi^{0\prime}_{2}\rangle=\frac{v_{u}}{\sqrt{2}},\;
 \langle\phi^{0\prime}_{3}\rangle=\cdots=\langle\phi^{0'}_{2n}\rangle=0 ,\;\;
 (v_{d},v_{u}:\mbox{real}), \label{pbasis}
\end{equation}
where prime denotes a quantity after the field transformation.

Just like in the MSSM case, we regard a diagonal element of the mass squared 
matrix, $\langle\phi |M^{2}|\phi\rangle$, as the desired upper bound.
Here the matrix $M^{2}$ is the mass squared matrix of the neutral Higgses, and 
the field~$\phi$ is the real part of the field~$\phi^{0}$ defined by
\begin{equation}
 \left(\begin{array}{c} \phi^{0} \\ \chi^{0} \end{array}\right)=
 \left(\begin{array}{cc} \cos\beta & \sin\beta \\ -\sin\beta & \cos\beta
 \end{array}\right)\left(\begin{array}{c} \phi^{0\prime}_{1} \\ 
 \phi^{0\prime}_{2} \end{array}\right),\;\;
 (\tan\beta\equiv\frac{v_{u}}{v_{d}})
\end{equation}
and 
\begin{equation}
 \phi^{0}=\frac{1}{\sqrt{2}}(\phi+i\varphi).
\end{equation}
Then we can restrict the effective potential to $\phi^{0\prime}_{1}$ and 
$\phi^{0\prime}_{2}$ directions.
\begin{eqnarray}
 V^{(0)}(\phi^{0\prime}_{1},\phi^{0\prime}_{2},\phi^{0\prime*}_{1},
 \phi^{0\prime*}_{2})&=&\frac{g^{2}+g'^{2}}{8}(|\phi^{0\prime}_{1}|^{2}
 -|\phi^{0\prime}_{2}|^{2})^{2} \nonumber \\
 &&+m_{1}^{\prime 2}|\phi^{0\prime}_{1}|^{2}
 +m_{2}^{\prime 2}|\phi^{0\prime}_{2}|^{2}
 -(m_{12}^{\prime 2}\phi^{0\prime}_{1}\phi^{0\prime}_{2}+h.c.).
\end{eqnarray}
This is the same form as the Higgs potential of the MSSM.
Therefore it is a trivial result that the desired tree-level upper bound 
$m_{h}$ is the same one as that of the MSSM, {\it i.e.}, 
$m_{h}=m_{Z}\cos\beta$ \cite{Flo;Sher}.

Next we consider the one-loop radiative corrections to this tree-level 
upper bound.
The one-loop corrections to the effective potential mainly come from the loops 
of quarks and squarks in the third generation because of the large sizes of 
their Yukawa couplings.
In the $\overline{\mbox{MS}}$ scheme, this is given by
\begin{equation}
 V^{(1 loop)}(\phi^{0}_{u},\phi^{0}_{d},\phi^{0*}_{u},\phi^{0*}_{d})=
 \frac{3}{32\pi^{2}}\left[\sum_{\tilde{q}=\tilde{t}_{1},\tilde{t}_{2},
 \tilde{b}_{1},\tilde{b}_{2}}m_{\tilde{q}}^{4}
 (\ln\frac{m_{\tilde{q}}^{2}}{Q^{2}}-\frac{3}{2})
 -2\sum_{q=t,b}m_{q}^{4}(\ln\frac{m_{q}^{2}}{Q^{2}}-\frac{3}{2})\right],
 \label{v1lp}
\end{equation}
where $m_{\tilde{q}}$ and $m_{q}$ are Higgs-field-dependent masses of the 
squarks and quarks respectively, and defined by
\begin{equation}
 \left.\begin{array}{c}m_{\tilde{t}_{1}}^{2} \\ m_{\tilde{t}_{2}}^{2} 
 \end{array}\right\}=m^{2}+m_{t}^{2}
 \pm |\mbox{}^{t}h_{t}A_{t}\phi^{0}_{u}
 +\mbox{}^{t}h_{t}\mu^{\dagger}\phi^{0*}_{d}|,
\end{equation}
\begin{equation}
 \left.\begin{array}{c}m_{\tilde{b}_{1}}^{2} \\ m_{\tilde{b}_{2}}^{2} 
 \end{array}\right\}=m^{2}+m_{b}^{2}
 \pm |\mbox{}^{t}h_{b}A_{b}\phi^{0}_{d}
 +\phi^{0\dagger}_{u}\mu^{\dagger}h_{b}|,
\end{equation}
\begin{eqnarray}
 m_{t}^{2}&=&|\mbox{}^{t}h_{t}\phi^{0}_{u}|^{2}, \\
 m_{b}^{2}&=&|\mbox{}^{t}h_{b}\phi^{0}_{d}|^{2}, 
\end{eqnarray}
\begin{equation}
 m_{Z}^{2}=\frac{1}{2}(g^{2}+g^{\prime 2})
 (|\phi^{0}_{d}|^{2}+|\phi^{0}_{u}|^{2})\;\;,\;\;\:
  \tan\beta=|\phi^{0}_{u}|/|\phi^{0}_{d}|.
\end{equation}
Here $h_{t}$ and $h_{b}$ are the row vectors of the Yukawa couplings defined by
\begin{equation}
 h_{t}\equiv\mbox{}^{t}(h_{2},h_{4},\cdots,h_{2n}) \hspace{5mm},\hspace{5mm}
 h_{b}\equiv\mbox{}^{t}(h_{1},h_{3},\cdots,h_{2n-1})\;,
\end{equation}
and $A_{t}={\rm diag}(A_{2},A_{4},\cdots,A_{2n})$, 
$A_{b}={\rm diag}(A_{1},A_{3},\cdots,A_{2n-1})$ are the soft SUSY-breaking 
$A$-parameter matrices, and $\mu$ is the $\mu$-parameter matrix
whose $(i,j)$-component is $\mu_{2i-1,2j}$.
Here we assume the soft SUSY-breaking squark masses to be degenerate and take
the common mass $m$.
We also neglect the small $D$-term contributions for simplicity, 
but including these contributions does not change the following discussion.

We should calculate the second derivative of 
$V^{(1)}\equiv V^{(0)}+V^{(1loop)}$
in terms of $\phi$ to obtain the upper bound $m_{h}$.
In the basis (\ref{pbasis}), the effective potential restricted to 
the directions of $\phi^{0\prime}_{1}$ and $\phi^{0\prime}_{2}$ again becomes 
the same form as that of the MSSM.
Hence we can easily see that the upper bound $m_{h}$ is also the same as 
the one in the MSSM case and conclude that $m_{h}=140\sim 150$~GeV, 
which is realized in the case of the large left-right stop mixing.

Thus $m_{h}$ is the same as the one in the MSSM up to one-loop order.
At two-loop order, however, the extra Yukawa couplings first contribute 
to $m_{h}$ as well as the strong gauge coupling~$\alpha_{s}$.
Then since two-loop order is the lowest order in terms of these couplings
and one-loop corrections are the same order as the tree-level value, 
the two-loop corrections might be relatively large.
In fact, the QCD correction lowers the upper bound by more than 10 GeV 
in the MSSM due to the large size of $\alpha_{s}$.
Since the extra Yukawa couplings are unconstrained experimentally, we will 
set their upper bounds by requiring that the theory remains perturbative up to 
some scale $\Lambda$.
The maximal values of the extra Yukawa couplings under this requirement 
might be large, so that 
the two-loop corrections to $m_{h}$ due to these couplings are as large as 
the QCD correction.
So in this letter, the upper bound $m_{h}$ is calculated using 
the RGE improved effective potential approach in the two-loop leading-log
approximation in order to see how different the upper bound $m_{h}$ is from
that of the MSSM.

\section{Upper bound formula}
For simplicity, we assume that the squarks are degenerate and their masses are 
1~TeV throughout this letter and neglect 
$O(g^{4},g^{2}g^{\prime 2},g^{\prime 4})$ contributions.
We can obtain the upper bound formula in a quite similar way to Ref.\cite{CQW},
in which the analytic expression of the lightest Higgs mass is provided in 
two-loop leading-log approximation in the MSSM.
\begin{eqnarray}
 m_{h}^{2}&=&\ovmz^{2}\cos^{2}2\beta\left\{1-\frac{3\ovmt^{2}}
 {2\pi^{2}v^{2}}\left[1-\frac{8\cos^{2}\beta}{\cos 2\beta}(1-\frac{\ovmb^{2}}
 {\ovmt^{2}}\tan^{2}\beta)\right]\ln\frac{m_{S}}{\ovmt}\right\} \nonumber \\
 &&+\frac{3\ovmt^{4}}{\pi^{2}v^{2}}\left\{\frac{1}{2}
 \frac{X_{t}^{2}}{m_{S}^{2}}
 (1-\frac{X_{t}^{2}}{12m_{S}^{2}})-\frac{\ovmz^{2}}{8\cos 2\beta\:\ovmt^{2}}
 \frac{X_{t}^{2}}{m_{S}^{2}}+\ln\frac{m_{S}}{\ovmt}\right. \nonumber \\
 &&\hspace{1.5cm}
 +\frac{1}{8\pi^{2}}\left.\left[(\beta_{m_{t}}[\ovmt]-6\frac{\ovmt^{2}}
 {v^{2}})\frac{X_{t}^{2}}{m_{S}^{2}}(1-\frac{X_{t}^{2}}{m_{S}^{2}})\ln
 \frac{m_{S}}{\ovmt}\right.\right. \nonumber \\
 &&\hspace{2.5cm}
 +\frac{\ovmz^{2}}{v^{2}\cos 2\beta}\left(6\cos^{2}\beta(1-\frac{\ovmb^{2}}
 {\ovmt^{2}}\tan^{2}\beta)-\frac{3}{2}\right)\frac{X_{t}^{2}}{m_{S}^{2}}
 \ln\frac{m_{S}}{\ovmt} \nonumber \\
 &&\hspace{2.5cm}
 +\left.\left.(\beta_{m_{t}}[\ovmt]-12\frac{\ovmt^{2}}{v^{2}})
 (\ln\frac{m_{S}}{\ovmt})^{2}\right]\right\} \label{upbd-bt}
\end{eqnarray}
where $\beta_{m_{t}}$ is defined as 
\begin{equation}
 Q\frac{\partial\;(h^{\prime}_{2}\sin\beta)}{\partial Q}\equiv
 \frac{h^{\prime}_{2}\sin\beta}{16\pi^{2}}\beta_{m_{t}}
\end{equation}
and $\beta_{m_{t}}[\ovmt]$ stands for
$\beta_{m_{t}}$ defined at the scale $Q=\ovmt$.
The renormalization scale is set to be $\ovmt$, and 
$\ovmz=m_{Z}(\ovmt),\;\ovmb=m_{b}(\ovmt)$.
The mass parameter $\ovmt$ is the on-shell running mass $m_{t}(M_{t})$, 
where $M_{t}$ is the pole mass, and the relation between $\ovmt$ and 
$M_{t}$ in the $\overline{\mbox{MS}}$ scheme is 
\begin{equation}
 \ovmt=\frac{M_{t}}{1+4\alpha_{s}(M_{t})/3\pi}.
\end{equation}
The parameter $X_{t}$ is the mixing parameter of the left-right stop mixing,
which is defined by $X_{t}\equiv |A^{\prime}_{2}+\mu^{\prime}_{12}\cot\beta|$,
and $m_{S}\equiv\sqrt{m^{2}+\ovmt^{2}}$ can be regarded as the SUSY-breaking 
scale.

The factor $\ovmz^{2}\cos^{2}2\beta$ in the first line of (\ref{upbd-bt}) 
corresponds to the tree-level contribution and the remaining factor 
of the first line represents the anomalous dimension of the lightest Higgs 
field between the SUSY-breaking scale $m_{S}$ and the renormalization scale
$\ovmt$.
The second line of the formula is the part of the one-loop correction, 
and the rest is the two-loop leading-log contributions. 

It is clear from the above expression that the upper bound $m_{h}$ 
gives the absolute upper bound when $\tan\beta$ is large and 
$\beta_{m_{t}}[\ovmt]$ has its maximal value.
The $\beta$-function $\beta_{m_{t}}$ is expressed 
by using the step functions as
\begin{eqnarray}
 \beta_{m_{t}}&=&\tilde{\beta}_{h^{\prime}_{2}}+\cos^{2}\beta
 (\gamma_{\phi^{\prime}_{2}}^{(1)}-\gamma_{\phi^{\prime}_{1}}^{(1)}) 
 \nonumber \\
 &=&-\frac{4}{3}g_{s}^{2}(6-\theta_{\tilde{G}\tilde{Q}}
 -\theta_{\tilde{G}\tilde{U}})+3h^{\prime 2}_{2}\sin^{2}\beta \nonumber \\
 &&+\frac{1}{2}\sum_{i=1}^{n}|h^{\prime}_{2i}|^{2}(3\theta_{\phi^{\prime}_{2i}}
 +2\theta_{\tilde{\phi}^{\prime}_{2i}\tilde{Q}}
 +\theta_{\tilde{\phi}^{\prime}_{2i}\tilde{U}})
 \nonumber \\
 &&+\frac{1}{2}\sum_{i=1}^{n}|h^{\prime}_{2i-1}|^{2}
 (\theta_{\phi^{\prime}_{2i-1}}
 +\theta_{\tilde{\phi}^{\prime}_{2i-1}\tilde{D}})+\cdots,\label{btmt}
\end{eqnarray}
where $g_{s}$ is the strong gauge coupling and the ellipsis denotes the terms 
involving $g,g^{\prime},\cdots$, which can be 
neglected because they are irrelevant within our approximations.
A step function $\theta_{P}$ is defined as one above the mass of the 
particle~$P$ and zero below it, and $\theta_{AB}\equiv\theta_{A}\theta_{B}$.
($\tilde{G}$ denotes gluino, $\tilde{Q}$, $\tilde{U}$ and $\tilde{D}$ are
the squarks in the third generation, $\phi^{\prime}_{i}$ are Higgses and 
$\tilde{\phi}^{\prime}_{i}$ are Higgsinos.)
Now since we consider below the scale 
$Q=m(=m_{\tilde{Q}}=m_{\tilde{U}}=m_{\tilde{D}})$,
the maximal value of $\beta_{m_{t}}$ is
\begin{equation}
 \beta_{m_{t}}=-8g_{s}^{2}+3h^{'2}_{2}\sin^{2}\beta+\frac{3}{2}\sum_{i=1}^{n}
 |h^{\prime}_{2i}|^{2}+\frac{1}{2}\sum_{i=1}^{n}|h^{\prime}_{2i-1}|^{2},
 \label{btmt-mx}
\end{equation}
which is realized in the limit that all the Higgs bosons are lighter than 
the top quark.

Here we rotate in the Higgs field space again and make the only 
one pair of the Higgs doublets couple to quarks.
The top and bottom Yukawa couplings in this basis are
\begin{equation}
 h^{''2}_{2}=\sum_{i=1}^{n}|h'_{2i}|^{2}\;,\hspace{5mm}
 h^{''2}_{1}=\sum_{i=1}^{n}|h'_{2i-1}|^{2}, \label{hpp}
\end{equation}
respectively.
(The quantities in this basis are denoted by the double primed ones.)

However, this limit does not coincide with the limit that the lightest mass 
eigenvalue saturates the upper bound $m_{h}$.
In fact, not all the Higgses have to be heavy to saturate $m_{h}$, 
but some of them may be relatively light 
unless they mix with the lightest mass eigenstate.
In this case the scale at which $\beta_{m_{t}}$ is estimated is not above 
all of the Higgs mass eigenvalues and thus $\beta_{m_{t}}[\ovmt]$ that should 
be used in the upper bound formula (\ref{upbd-bt}) is smaller than the maximal 
value (\ref{btmt-mx}) 
by the contributions of the ``extra'' Yukawa couplings corresponding to 
the ``extra'' Higgses that are decoupled.
Here we denote the word ``extra'' as the meaning of giving no masses to the 
quarks.
For example, the extra Yukawa couplings mean $h^{\prime}_{2i}$ and 
$h^{\prime}_{2i-1}$ ($i\geq 2$).
The absolute upper bound is thus realized when 
all Higgses that do not mix with the lightest one are light and 
the extra Yukawa couplings that correspond to heavy extra Higgses are small 
enough so that $\beta_{m_{t}}$ is close to (\ref{btmt-mx})
if the fixed values of $h''_{2}$ and $h''_{1}$ are given.
Thus, from now on, we will restrict the parameter space to this region.

Unlike the MSSM case, there is no upper bound on $h''_{2}$ and $h''_{1}$
due to the existence of unconstrained coupling constants
$h'_{2i}$ and $h'_{2i-1}$ $(i\geq 2)$.
Then we put the requirement : 
\begin{center}
\bf ``The theory remains perturbative up to some scale $\Lambda$.''
\end{center}

By substituting (\ref{btmt-mx}) for $\beta_{m_{t}}[\ovmt]$ 
in the formula (\ref{upbd-bt})and setting $\tan\beta$ large, 
the desired upper bound is obtained as
\begin{eqnarray}
 m_{h}^{2}=&&\ovmz^{2}\left(1-\frac{3\ovmt^{2}}{2\pi^{2}v^{2}}
 \ln\frac{m_{S}}{\ovmt}\right) \nonumber \\
 &&+\frac{3\ovmt^{4}}{\pi^{2}v^{2}}\left\{\frac{1}{2}
 \frac{X_{t}^{2}}{m_{S}^{2}}
 (1-\frac{X_{t}^{2}}{12m_{S}^{2}})+\frac{\ovmz^{2}}{8\ovmt^{2}}
 \frac{X_{t}^{2}}{m_{S}^{2}}+\ln\frac{m_{S}}{\ovmt}\right.\nonumber \\
 &&\hspace{1.5cm}+\frac{1}{8\pi^{2}}\left[(\frac{3}{2}h_{2\mx}^{''2}
 +\frac{1}{2}h_{1\mx}^{''2}
 -32\pi\alpha_{s})\frac{X_{t}^{2}}{m_{S}^{2}}(1-\frac{X_{t}^{2}}{12m_{S}^{2}})
 \ln\frac{m_{S}}{\ovmt}\right. \nonumber \\
 &&\hspace{2.5cm}-\frac{9\ovmz^{2}}{2v^{2}}\frac{X_{t}^{2}}{m_{S}^{2}}
 \ln\frac{m_{S}}{\ovmt} \nonumber \\
 &&\hspace{2.5cm}+\left.\left.(-6\frac{\ovmt^{2}}{v^{2}}
 +\frac{3}{2}h_{2\mx}^{''2}+\frac{1}{2}h_{1\mx}^{''2}-32\pi\alpha_{s})
 (\ln\frac{m_{S}}{\ovmt})^{2}
 \right]\right\},\nonumber \\
\end{eqnarray}
where $h''_{2\mx}$ and $h''_{1\mx}$ are the values of $h''_{2}$ and $h''_{1}$ 
that maximize the combination $\frac{3}{2}h^{''2}_{2}+\frac{1}{2}h^{''2}_{1}$ 
under the above requirement.

\section{Maximal values of $h''_{2}$ and $h''_{1}$}
Now we will compute $h''_{2\mx}$ and $h''_{1\mx}$.
First, we should notice that the RGEs of $h''_{2}$ and $h''_{1}$ are the same
as that of $h_{t}$ and $h_{b}$ in the MSSM because  
$h''_{2}$ and $h''_{1}$ are the sole pair of the Yukawa couplings that 
couple to the up-type quark multiplets and the down-type quark multiplets 
respectively and the gauge interactions do not mix $\phi^{''}_{1}$ and 
$\phi^{''}_{2}$ with the other Higgses. 
Thus, we can use the formulae in \cite{Lahanas}, which are expressed by the 
step functions just like (\ref{btmt}).
Here we will use the RGEs that is gained by setting 
all $\theta$'s to one above the SUSY-breaking scale $m_{S}$ 
and all $\theta$'s except $\theta_{\tilde{Q}}$, $\theta_{\tilde{U}}$ 
and $\theta_{\tilde{D}}$ to one below the scale $m_{S}$ at the RGE formulae
in \cite{Lahanas}.
Of course the correct RGEs below $m_{S}$ depend on the mass spectrum 
of the SUSY particles, but this difference is thought to be small and 
negligible.
Further we will also neglect the $\tau$ Yukawa coupling $h''_{\tau}$ 
for simplicity.
Neglecting $h''_{\tau}$ does not lower the upper bound $m_{h}$ because
the coefficient of $h^{''2}_{\tau}$ in the RGE of $h''_{1}$ is positive.

Thus the RGEs we are using are
\begin{eqnarray}
 Q\frac{\partial h''_{2}}{\partial Q}&=&\frac{h''_{2}}{16\pi^{2}}\left(
 -\frac{16}{3}g_{3}^{2}-3g_{2}^{2}-\frac{13}{15}g_{1}^{2}
 +6h^{''2}_{2}+h^{''2}_{1}\right), \label{RGE-h2pp} \\
 Q\frac{\partial h''_{1}}{\partial Q}&=&\frac{h''_{1}}{16\pi^{2}}\left(
 -\frac{16}{3}g_{3}^{2}-3g_{2}^{2}-\frac{7}{15}g_{1}^{2}
 +h^{''2}_{2}+6h^{''2}_{1}\right), \label{RGE-h1pp}
\end{eqnarray}
above $m_{S}$ and
\begin{eqnarray}
 Q\frac{\partial h''_{2}}{\partial Q}&=&\frac{h''_{2}}{16\pi^{2}}\left(
 -8g_{3}^{2}-\frac{3}{4}g_{2}^{2}-\frac{11}{20}g_{1}^{2}
 +\frac{9}{2}h^{''2}_{2}+\frac{1}{2}h^{''2}_{1}\right), \label{RGE-h2pp-bl} \\
 Q\frac{\partial h''_{1}}{\partial Q}&=&\frac{h''_{1}}{16\pi^{2}}\left(
 -8g_{3}^{2}-\frac{3}{4}g_{2}^{2}+\frac{1}{20}g_{1}^{2}
 +\frac{1}{2}h^{''2}_{2}+\frac{9}{2}h^{''2}_{1}\right), \label{RGE-h1pp-bl}
\end{eqnarray}
below $m_{S}$.
Here $g_{3}=g_{s}$, $g_{2}=g$ and $g_{1}=\sqrt{5/3}g'$.

Next we will consider the gauge couplings.
By extending the RGEs in \cite{Lahanas} in the same way as the above 
Yukawa coupling case and solving the RGEs, we obtain above the scale $m_{S}$,
\begin{eqnarray}
 g_{3}^{2}&=&\frac{1}{\frac{3}{8\pi^{2}}\ln(Q/M_{3})}, \label{RGE-g3} \\
 g_{2}^{2}&=&\frac{1}{-\frac{n}{8\pi^{2}}\ln(Q/M_{2})}, \label{RGE-g2} \\
 g_{1}^{2}&=&\frac{1}{-\frac{30+3n}{40\pi^{2}}\ln(Q/M_{1})},\label{RGE-g1}
\end{eqnarray}
and below $m_{S}$,
\begin{eqnarray}
 g_{3}^{2}&=&\frac{1}{\frac{5}{8\pi^{2}}\ln(Q/M'_{3})}, \label{RGE-g3-bl} \\
 g_{2}^{2}&=&\left\{\begin{array}{l} {\displaystyle 
 \frac{1}{-\frac{n-2}{8\pi^{2}}\ln(Q/M'_{2})} \hspace{5mm} 
 (\mbox{for}\;\;n\neq 2)} \\
 {\mbox{constant} \hspace{1.8cm} (\mbox{for}\;\;n=2)} \end{array}\right.\;\;, 
 \label{RGE-g2-bl} \\
 g_{1}^{2}&=&\frac{1}{-\frac{49+6n}{80\pi^{2}}\ln(Q/M'_{1})},\label{RGE-g1-bl}
\end{eqnarray}
where $M_{i}$ and $M'_{i}$ ($i=1,2,3$) are decided by the initial conditions: 
\begin{equation}
 \alpha_{3}(\ovmz)=0.12,\hspace{3mm} \alpha_{2}(\ovmz)=0.034,\hspace{3mm} 
 \alpha_{1}(\ovmz)=0.017, \label{in-cd:gauge}
\end{equation}
and the boundary conditions at $Q=m_{S}$.

Using (\ref{RGE-h2pp})-(\ref{RGE-g1-bl}), 
$h''_{2\mx}$ and $h''_{1\mx}$ can be computed.

As the criterion that the theory remains perturbative up to $\Lambda$, 
we adopt the condition whether one of the dimensionless coupling constants 
(gauge or Yukawa) saturates the scale $\Lambda$.
A particular coupling $\lambda$ is said to ``saturate'' a scale $\Lambda$ if
\begin{equation}
 \frac{\lambda^{2}(Q^{2})}{4\pi}\leq 1 \label{saturate}
\end{equation}
for $Q\leq\Lambda$, and the equality in (\ref{saturate}) holds for $Q=\Lambda$.
One can see from the RGEs (\ref{RGE-g3})-(\ref{RGE-g1-bl}) and the conditions
(\ref{in-cd:gauge}) that $g_{2}$ is the first gauge coupling to saturate
for every scale $\Lambda$.
Thus one cannot set $\Lambda$ larger than the scale $\Lambda_{n}$ determined
by the saturation of $g_{2}$ for a given $n$.
However, this constraint does not restrict the allowed region of $\Lambda$
so strictly.
For example, $\Lambda_{2}=10^{42.2}$~GeV, $\Lambda_{5}=10^{18.0}$~GeV, 
$\Lambda_{6}=10^{15.4}$~GeV and $\Lambda_{10}=10^{10.0}$~GeV.

\section{Self-energy contribution}
The upper bound $m_{h}^{2}$ discussed so far is 
$\partial^{2}V^{(1)}/\partial\phi^{2}$.
Since the effective potential~$V$ is a sum of one-particle-irreducible (1PI)
Feynman diagrams with zero external momenta, $m_{h}^{2}$ is written as
\begin{equation}
 m_{h}^{2}=m_{h\mbox{\scriptsize tree}}^{2}+\Pi(0), \label{runmh}
\end{equation}
where $\Pi(p^{2})$ is the self-energy of the Higgs boson.

On the other hand, physical mass $M_{h}$ is defined as a pole of
the propagator, that is,
\begin{equation}
 M_{h}^{2}=m_{h\mbox{\scriptsize tree}}^{2}+\Pi(M_{h}^{2}). \label{polemh}
\end{equation}

Then from (\ref{runmh}) and (\ref{polemh}), the relation between $m_{h}$
and $M_{h}$ is
\begin{equation}
 M_{h}^{2}=m_{h}^{2}+\Pi(M_{h}^{2})-\Pi(0). \label{run-pole}
\end{equation}
More precisely,
\begin{equation}
 M_{h}^{2}=m_{h}^{2}+{\rm Re}\Pi(M_{h}^{2})-{\rm Re}\Pi(0).
\end{equation}
The imaginary part of $\Pi(M_{h}^{2})-\Pi(0)$ contributes to the 
decay width of the Higgs boson, and if we suppose $M_{h}<2M_{W}$, where
$M_{W}$ is the physical mass of the W boson, the Higgs is stable
at tree level and (\ref{run-pole}) is correct.
   
This correction $\Delta\Pi(M_{h}^{2})\equiv\Pi(M_{h}^{2})-\Pi(0)$ 
mainly comes from one-loop electroweak interaction effects.

This correction is larger than the MSSM case due to the existence of the extra
Yukawa couplings, but still small and lift up the upper bound $m_{h}$ by 2~GeV
at most.

\section{Results}
\begin{figure}[p]
 \leavevmode
 \epsfysize=8cm
 \epsfxsize=8cm
 \centerline{\epsfbox{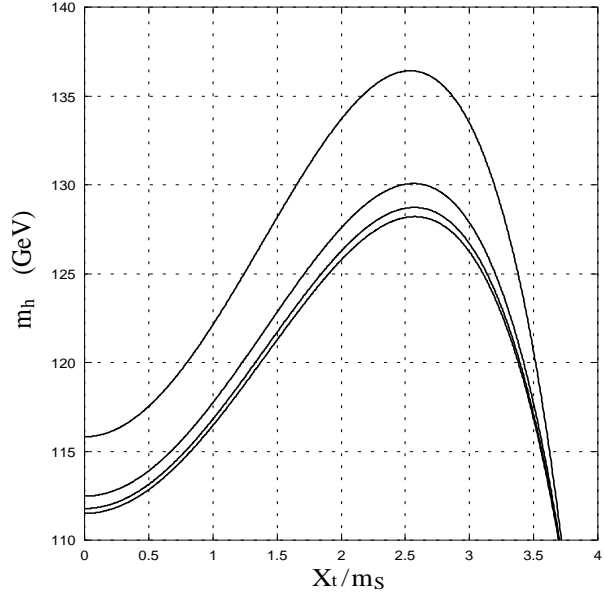}}
 \caption{The upper bound $m_{h}$ in the case of $n=2$.The lines corresponds
 to the case of $\Lambda=10^{16},10^{12},10^{8},10^{4}$ GeV from 
 bottom to top. The parameters, $\tan\beta$ and the common soft SUSY-breaking
 stop mass $m$, are set to be 40 and 1~TeV respectively.}
 \label{md1}
\end{figure}

The results including the self-energy contribution is plotted 
in Fig.\ref{md1} in the case of $n=2$.
Since we are interested in the absolute upper bound on the mass of 
the lightest Higgs, only large $\tan\beta$ case ($\tan\beta=40$) is plotted.
The lines are the case of $\Lambda=10^{16},10^{12},10^{8},10^{4}$~GeV from
bottom to top.
The common soft SUSY-breaking stop mass parameter $m$ is set to be 1~TeV
in all the cases.
We can see from this figure that setting the saturation scale $\Lambda$ 
larger, the upper bound $m_{h}$ becomes more stringent.

\begin{figure}
 \leavevmode
 \epsfysize=8cm
 \epsfxsize=8cm
 \centerline{\epsfbox{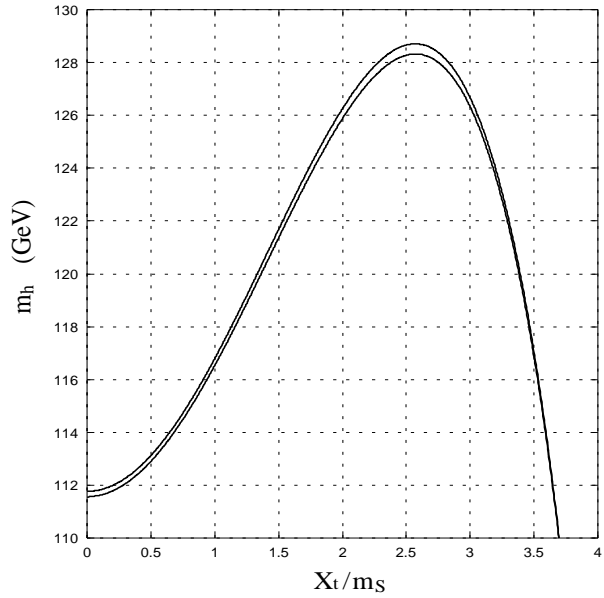}}
 \caption{The comparison of the two cases: $n=2$ and $n=6$ 
 for $\Lambda=10^{15}$ GeV. In both cases we set $\tan\beta=40$ and $m=1$~TeV.}
 \label{md2}
\end{figure}

Because the RGEs (\ref{RGE-h2pp})-(\ref{RGE-h1pp-bl}) have $n$-dependence 
only through the terms involving the weak gauge couplings $g$ and $g'$,
the $n$-dependence of $m_{h}$ is quite small.
In fact, since the $n$-dependence of the result becomes larger as 
$\Lambda$ is larger and $g$ becomes no longer small at 
$10^{15}$~GeV for $n=6$ ($\Lambda_{6}=10^{15.4}$~GeV),  
we can see the most enhanced $n$-dependence of $m_{h}$  
when we compare the two cases: $n=2$ and $n=6$ for $\Lambda=10^{15}$~GeV.
This situation is plotted in Fig.\ref{md2}.

From this figure we can see that the $n$-dependence of $m_{h}$ can be
neglected.

\vspace{1cm}
\mbox{}\linebreak
{\Large\bf Acknowledgements}

\mbox{}\linebreak
The author would like to thank N.Sakai for useful advice and 
careful checking of my idea and calculations in spite of his hard schedule, 
and Y.Okada for an important suggestion.
I would also like to thank M.Hayakawa, S.Kiyoura and Y.Chikira
for helpful discussion and valuable information.


\begin{thebibliography}{100} 
 \bibitem{Ino;Kaku} K.Inoue, A.Kakuto, H.Komatsu and S.Takeshita, 
{\it Prog.Theor.Phys.} {\bf 67} (1982) 1889.

 \bibitem{Okada;Yama} Y.Okada, M.Yamaguchi and T.Yanagida, 
 {\it Prog.Theor.Phys.} {\bf 85} (1991) 1.
 
 \bibitem{Ellis;R;Z} J.Ellis, G.Ridolfi and F.Zwirner, {\it Phys.Lett.} 
 {\bf B262} (1991) 477.

 \bibitem{Haber;Hemp} H.Haber and R.Hempfling, {\it Phys.Rev.Lett.} {\bf 66} 
 (1991) 1815.

 \bibitem{Okada;Yama-2} Y.Okada, M.Yamaguchi and T.Yanagida, {\it Phys.Lett.} 
{\bf B262} (1991) 54.

 \bibitem{Kodaira} J.Kodaira, Y.Yasui and K.Sasaki, {\it Phys.Rev.} {\bf D50} 
 (1994) 7035.

 \bibitem{Hemp;Hoang} R.Hempfling and A.H.Hoang, {\it Phys.Lett.} {\bf B331}
 (1994) 99.

 \bibitem{Casas-2} J.A.Casas, J.R.Espinosa, M.Quir\'{o}s and A.Riotto, 
 {\it Nucl.Phys.} {\bf B436} (1995) 3; (Erratum) {\bf B439} (1995) 466.

 \bibitem{3H} H.E.Haber, R.Hempfling and A.H.Hoang, {\it Z.Phys.} {\bf C75}
 (1997) 539.

 \bibitem{Hein;Holl} S.Heinemeyer, W.Hollik and G.Weiglein, 
 {\tt hep-ph/9803277};\\ {\tt hep-ph/9807423}.

 \bibitem{Zhang} R.Zhang, {\tt hep-ph/9808299}.

 \bibitem{Flo;Sher} R.Flores and M.Sher, {\it Ann.of Phys.} {\bf 148} (1983) 
 95.

 \bibitem{CQW} M.Carena, M.Quir\'{o}s and C.E.M.Wagner, {\it Nucl.Phys.}
  {\bf B461} (1996) 407.

 \bibitem{Lahanas} A.B.Lahanas and K.Tamvakis, {\it Phys.Lett.} {\bf B348} 
 (1995) 451.
\end{thebibliography}
\end{document}